\documentclass[prd,nopacs,preprintnumbers,twocolumn,amsmath,nofootinbib,amssymb]{revtex4}
\usepackage{graphicx,color,dcolumn,booktabs,bm}
\usepackage{longtable,lscape}
\usepackage{txfonts}
\usepackage{overpic}
\usepackage{epstopdf}
\usepackage{indentfirst}
\usepackage{feynmf}   %{feynmp}
\usepackage{slashed}  %for Feynman symbols
\usepackage{cases}
\usepackage{color}
\usepackage{float}
\usepackage{gensymb}
\usepackage{multirow}
\usepackage{epsfig,dsfont,amssymb,amsmath,amsfonts,amsbsy,mathrsfs}
\graphicspath{{}} %

\usepackage{hyperref}
\hypersetup{colorlinks,citecolor=blue,anchorcolor=red,menucolor=red,linkcolor=red,filecolor=red,runcolor=red,urlcolor=blue,frenchlinks=red}

\makeatletter
\@addtoreset{equation}{section}
\makeatother

\allowdisplaybreaks

\newcolumntype{I}{!{\vrule width 0.9pt}}

\begin{document}

\title{Investigating the $\mathbf{\Upsilon(10753)\to \Upsilon(1^3D_J)\eta}$ transitions}
\author{Yu-Shuai Li$^{1,2}$}\email{liysh20@lzu.edu.cn}
\author{Zi-Yue Bai$^{1,2}$}\email{baizy15@lzu.edu.cn}
\author{Xiang Liu$^{1,2,3}$\footnote{Corresponding author}}\email{xiangliu@lzu.edu.cn}
\affiliation{$^1$School of Physical Science and Technology, Lanzhou University, Lanzhou 730000, China\\
$^2$Research Center for Hadron and CSR Physics, Lanzhou University and Institute of Modern Physics of CAS, Lanzhou 730000, China\\
$^3$Lanzhou Center for Theoretical Physics and Frontier Science Center for Rare Isotopes, Lanzhou University, Lanzhou 730000, China}

\begin{abstract}
In this work, we investigate the $\Upsilon(10753)\to\Upsilon(1^3D_J)\eta$ ($J=1,2,3$) processes, where the $\Upsilon(10753)$ is assigned as a conventional bottomonium under the $4S$-$3D$ mixing scheme. Our result shows that the concerned processes have considerable branching ratios, {\it i.e.},
branching ratios $\mathcal{B}[\Upsilon(10753)\to\Upsilon(1^3D_{1})\eta]$ and $\mathcal{B}[\Upsilon(10753)\to\Upsilon(1^3D_{2})\eta]$ can reach up to the order of magnitude of $10^{-4}-10^{-3}$, while $\mathcal{B}[\Upsilon(10753)\to\Upsilon(1^3D_{3})\eta]$ is around $10^{-6}-10^{-5}$. With the running of Belle II, it is a good opportunity for finding out the concerned hidden-bottom hadronic decays.
\end{abstract}

\maketitle

\section{introduction}
\label{sec01}

As presented in the Belle II physics book~\cite{Belle-II:2018jsg}, the designed luminosity of SuperKEKB can reach up to $8\times10^{35}\text{cm}^{-2}\text{s}^{-1}$. Thus, the forthcoming Belle II experiment represents the precision  frontier of particle physics, which is an ideal platform to perform the correlative study around heavy flavor physics. Obviously, some higher bottomonia can be accessible at Belle II, which may provide valuable hints to construct the bottomonium family.

Recently, the $\Upsilon(10753)$ was reported by the Belle Collaboration by analyzing the $e^+e^-\to\Upsilon(nS)\pi^+\pi^-$ ($n=1,2,3$) processes~\cite{Belle:2019cbt}. As a vector state, the $\Upsilon(10753)$ was also collected in Particle Data Group (PDG)~\cite{ParticleDataGroup:2020ssz}. Since a peculiar property of the $\Upsilon(10753)$ is its mass lower than the predicted mass of the $\Upsilon(3^3D_1)$ by the quenched potential model~\cite{Godfrey:2015dia,Segovia:2016xqb,Wang:2018rjg}, thus different theoretical groups tried to explain the observed $\Upsilon(10753)$ as exotic states like the tetraquark state~\cite{Wang:2019veq,Ali:2020svd}, the hybrid state~\cite{TarrusCastella:2021pld}, and the kinetic effect~\cite{Bicudo:2019ymo,Bicudo:2020qhp}.

\begin{figure}[htbp]\centering
  \includegraphics[width=76mm]{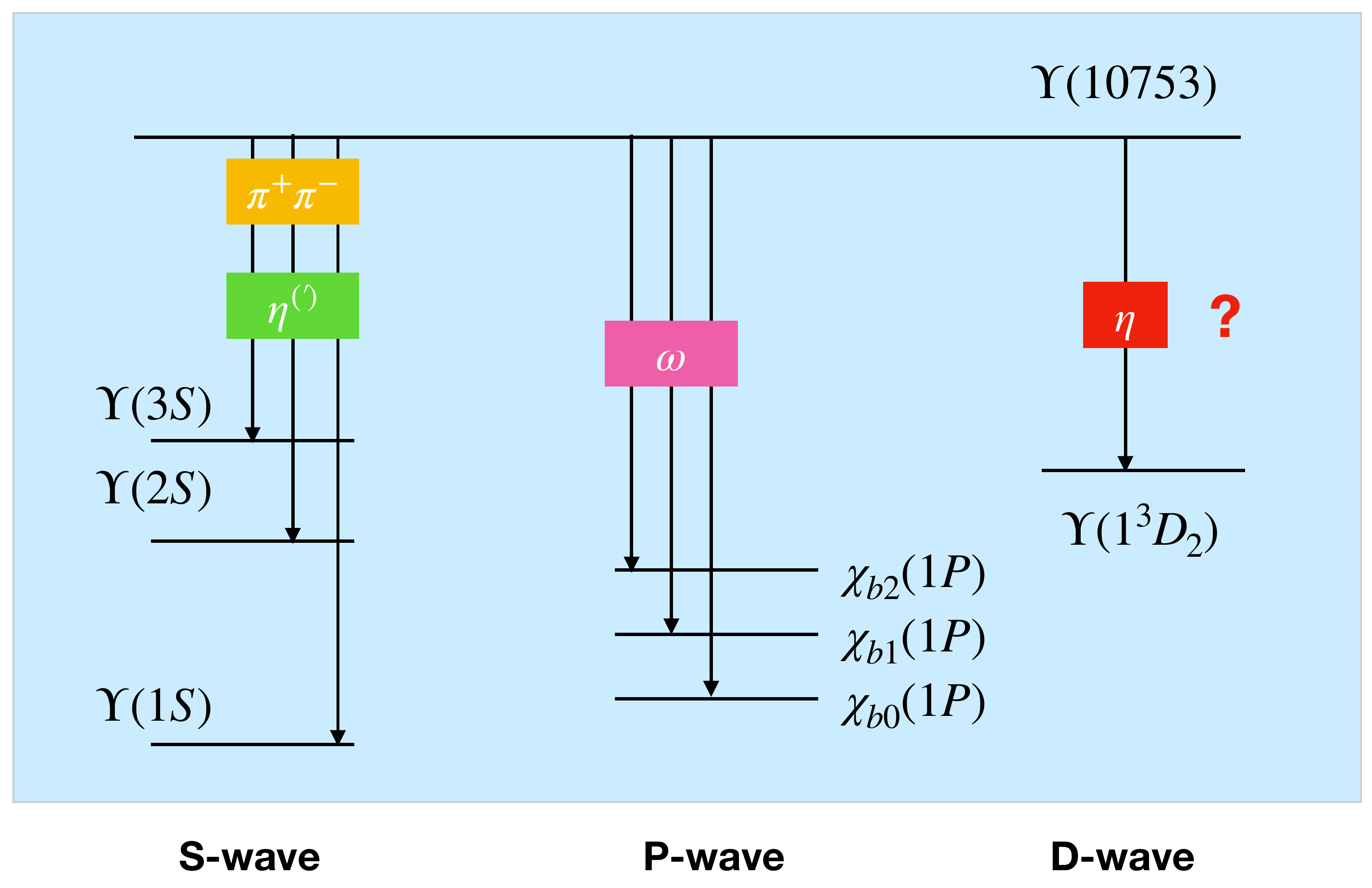}\\
  \caption{The present status of the study on the transitions of the $\Upsilon(10753)$ into other bottomonia.}
\label{fig:figsec01}
\end{figure}

Inspired by the $2S$-$1D$ mixing scheme for the charmonium $\psi(3770)$, we have a reason to believe that the $S$-$D$ mixing scheme should be considered when revealing the nature of the $\Upsilon(10753)$. Thus, the Lanzhou group introduced the $4S$-$3D$ mixing scheme to clarify the puzzling phenomenon of the $\Upsilon(10753)$~\cite{Li:2021jjt,Bai:2022cfz}.
{In Ref.~\cite{Li:2021jjt}, the Lanzhou group proposed the $4S$-$3D$ mixing scheme, which can solve the mass puzzle of the $\Upsilon(10753)$. A later result in Ref.~\cite{Bai:2022cfz} shows that the $\Upsilon(10753)$
under this mixing scheme has sizable dielectron decay width and   the measured values $\mathcal{R}_{n}=\Gamma_{e^+e^-}\times\mathcal{B}[\Upsilon(10753)\to\Upsilon(nS)\pi^+\pi^-]$ ($n=1,2,3$) by Belle~\cite{Belle:2019cbt} can be reproduced. In  summary, the current measured data of the $\Upsilon(10753)$~\cite{Belle:2019cbt}, including its mass and $\mathcal{R}_{n}$ values, can well be understood under the $4S$-$3D$ mixing scheme. Thus, the $\Upsilon(10753)$ can be still a good candidate of vector bottomonium.}
Along this line, several typical transitions of the $\Upsilon(10753)$ into other bottomonia with lower mass were explored in Refs.~\cite{Li:2021jjt,Bai:2022cfz}, which will be accessible at a future experiment like Belle II. In Fig.~\ref{fig:figsec01}, we summarize the present status of the study of the transitions of the $\Upsilon(10753)$ into other bottomonia.

Obviously, our knowledge of the transitions of the $\Upsilon(10753)$ into other bottomonia is still absent. A typical example is that the allowed $\Upsilon(10753)\to \Upsilon(1^3D_2)\eta$ is waiting to be  explored, not only by theorists but also by experimentalists. This fact stimulates our interest in carrying out the investigation of $\Upsilon(10753)\to \Upsilon(1^3D_J)\eta$ ($J=1,2,3$), where the $\Upsilon(1^3D_J)$ denote three $1D$ bottomonium states. By checking the PDG values~\cite{ParticleDataGroup:2020ssz},  we may find that only $\Upsilon(1^3D_2)$ was observed. For the remaining $1D$ bottomonia, they are still missing in experiment. Thus, the present study of $\Upsilon(10753)\to \Upsilon(1^3D_J)\eta$ has a close relation to these two missing bottomonia $\Upsilon(1^3D_1)$ and $\Upsilon(1^3D_3)$.

For calculating the branching ratio of the $\Upsilon(10753)\to \Upsilon(1^3D_J)\eta$ transitions, the concrete phenomenological model should be involved. Borrowing the former experience of the decays of higher states of heavy quarkonium, the coupled channel effect should be considered here~\cite{Meng:2007tk,Meng:2008dd,Meng:2008bq,Chen:2011jp,Chen:2011qx,Chen:2011zv,Chen:2014ccr,Wang:2016qmz,
Huang:2017kkg,Zhang:2018eeo,Huang:2018pmk,Huang:2018cco,Li:2021jjt,Bai:2022cfz}. In this work, we adopt the hadronic loop mechanism to present the concrete calculation, which will be mentioned in the following section.
We hope that our realistic investigation of the discussed processes may provide valuable information to experimentally search for $\Upsilon(10753)\to \Upsilon(1^3D_J)\eta$, which will be an intriguing research task for Belle II.

This paper is organized as follows. After the introduction, we illustrate the detailed calculation of $\Upsilon(10753)\to\Upsilon(1^3D_J)\eta$ ($J=1,2,3$) with the hadronic loop mechanism in Sec.~\ref{sec02}. And then, the numerical results are presented in Sec.~\ref{sec03}. Finally, we end the paper with a discussion and conclusion.

\section{the $\mathbf{\Upsilon(10753)\to\Upsilon(1^3D_J)\eta}$ transitions via the hadronic loop mechanism}
\label{sec02}

{Before studying the decays, we need to briefly introduce the $4S$-$3D$ mixing scheme. If assigning the $\Upsilon(10753)$ as the pure $\Upsilon(3D)$ state, the predicted mass of pure $\Upsilon(3D)$ state ranges from $10653$ MeV to $10717$ MeV~\cite{Badalian:2008ik,Badalian:2009bu,Godfrey:2015dia,Segovia:2016xqb,Wang:2018rjg}. Thus, there exists difference between the theoretical result and current measurement of the $\Upsilon(10753)$. Furthermore, the dielectron width of the $\Upsilon(3D)$ was estimated to be just a few eV~\cite{Godfrey:2015dia,Wang:2018rjg,Badalian:2008ik,Badalian:2009bu}, which is lower than the corresponding dielectron widths of the $\Upsilon(4S)$ and $\Upsilon(5S)$ states. Thus, it is difficult to find pure $\Upsilon(3D)$ state via the electron-positron annihilation process. However, the $\Upsilon(10753)$ signal was observed in the $e^{+}e^{-}\to\Upsilon(nS)\pi^{+}\pi^{-}$ processes by Belle~\cite{Belle:2019cbt}, which is puzzling for us. 
As proposed in Refs.~\cite{Li:2021jjt,Bai:2022cfz}, the $4S$-$3D$ mixing scheme for the $\Upsilon(10753)$ was introduced 
\begin{equation}
\left(\begin{array}{c}
\Upsilon_{4S-3D}\\
\Upsilon^{\prime}_{4S-3D}
\end{array}\right)
=\left(\begin{array}{cc}
\cos{\theta} & -\sin{\theta}\\
\sin{\theta} & \cos{\theta}
\end{array}\right)
\left(\begin{array}{c}
\Upsilon({4S})\\
\Upsilon({3D})
\end{array}\right),
\end{equation}
where $\theta$ denotes the mixing angle, and $\Upsilon_{{4S-3D}}$ and $\Upsilon_{{4S-3D}}^\prime$ are physical states. Here, the  $\Upsilon_{{4S-3D}}^\prime$ state corresponds to the observed $\Upsilon(10753)$. Obviously, the puzzle on mass can be solved as shown in Fig.~1 of Ref.~\cite{Li:2021jjt}, and the dielectron decay width of the $\Upsilon(10753)$ is sizable. Thus, the $\Upsilon(10753)$ still can be as a good candidate of vector bottomonium.}

\begin{figure}[htbp]\centering
  \includegraphics[width=80mm]{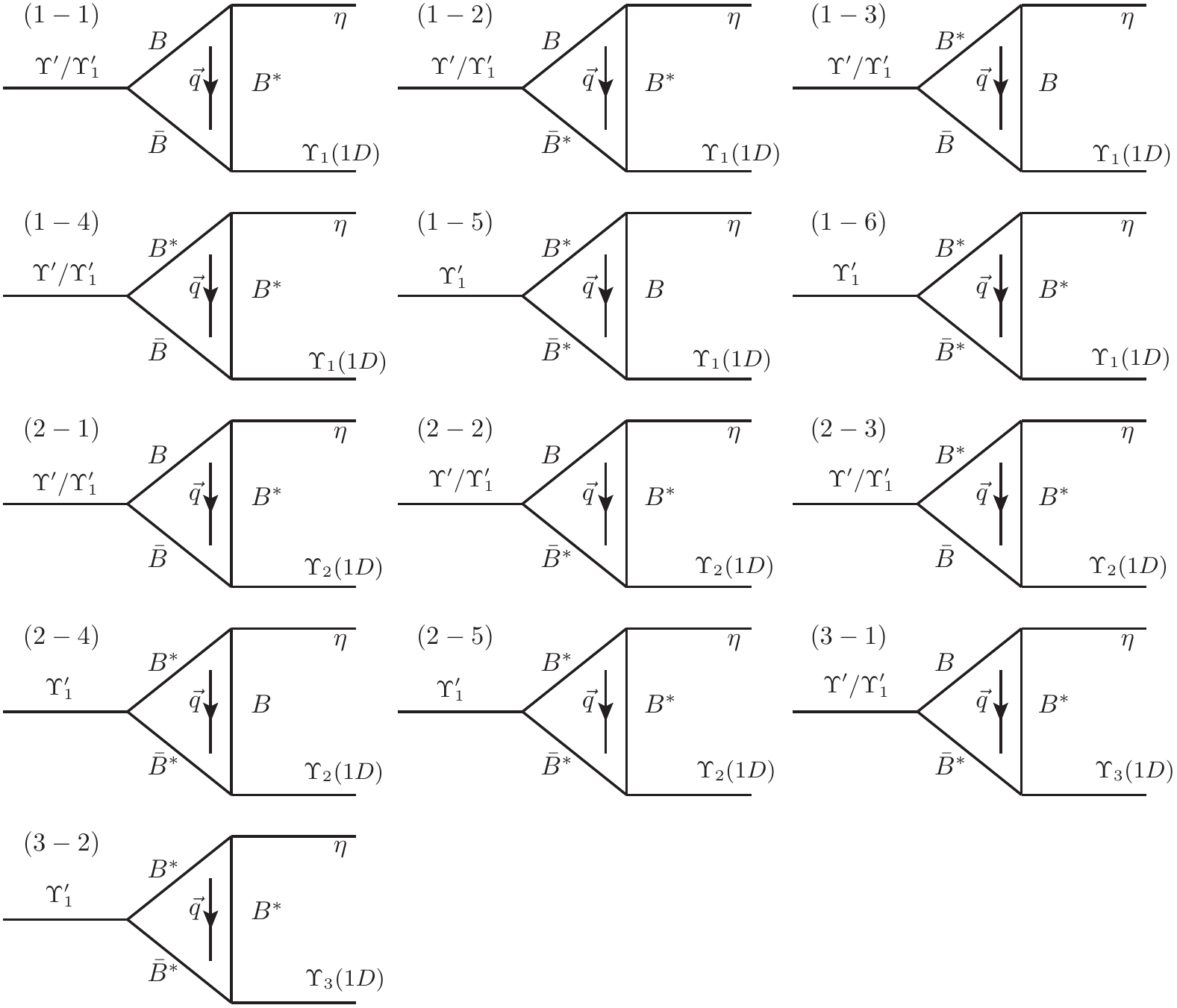}\\
  \caption{The schematic diagrams for depicting the $\Upsilon(4S,3D)\to\Upsilon(1^3D_J)\eta$ decays via the hadronic loop mechanism, where $\Upsilon^\prime$ and $\Upsilon_1^\prime$ represent the $\Upsilon(4S)$ and $\Upsilon(3D)$ components of the $\Upsilon(10753)$, respectively.}
\label{fig:DJeta}
\end{figure}

Based on hadronic loop mechanism, the initial $\Upsilon(10753)$ can be converted into final low-lying $D$-wave bottomonium $\Upsilon(1^3D_J)$ through the triangle loops composed of bottom mesons. The concerned diagrams are displayed in Fig.~\ref{fig:DJeta}, where the contributions from the $B_s^{(*)}$ meson loops can be ignored due to the weak coupling between the $\Upsilon(10753)$ and the $B_s^{(*)}\bar{B}_s^{(*)}$ pair~\cite{Liang:2019geg}.

For the diagrams shown in Fig.~\ref{fig:DJeta}, the general expression of their amplitude mediated by the hadronic loop mechanism reads as
\begin{equation}
\mathcal{M}=\int\frac{d^{4}q}{(2\pi)^{4}}\frac{\mathcal{V}_{1}\mathcal{V}_{2}\mathcal{V}_{3}}{\mathcal{P}_{1}\mathcal{P}_{2}\mathcal{P}_{E}}\mathcal{F}^{2}(q^2,m_{E}^2),
\label{eq:amplitude}
\end{equation}
where $\mathcal{V}_{\text{i}}$ ($i=1,2,3$) are interaction vertices, and $1/\mathcal{P}_{1,2,E}$ denote the corresponding propagators of intermediate bottom mesons. In addition, the form factor $\mathcal{F}(q^{2},m_{E}^2)$ should be introduced to compensate the off shell effect of the exchanged $B^{(*)}$ meson and depict the structure effect of interaction vertices~\cite{Locher:1993cc,Li:1996yn,Cheng:2004ru}. In our calculation, the monopole form factor~\cite{Gortchakov:1995im}
\begin{equation}
\mathcal{F}(q^2,m_{E}^2)=\frac{\Lambda^2-m_{E}^2}{\Lambda^2-q^2}\quad {\text with}\quad
\Lambda=m_{E}+\alpha_{\Lambda}\Lambda_{QCD}
\label{eq:formfactor}
\end{equation}
emphasized by QCD sum rules is adopted with $m_{E}$ and $q$ denoting the mass and four-momentum of the exchanged intermediate meson, respectively. Here, we take $\Lambda_{QCD}=220$ MeV~\cite{Liu:2006dq,Liu:2009dr,Li:2013zcr}, and $\alpha_{\Lambda}$ is a phenomenological dimensionless parameter.

The effective Lagrangian approach is used to give the concrete expressions of the decay amplitudes defined in Eq.~\eqref{eq:amplitude}. Due to the requirement from the heavy quark limit and the chiral symmetry, the concerned effective Lagrangians include~\cite{Casalbuoni:1996pg,Wise:1992hn,Xu:2016kbn,Duan:2021bna}
\begin{equation}
\begin{split}
\mathcal{L}_{S}=&\ ig_{S}\text{Tr}[S^{(b\bar{b})}\bar{H}^{(\bar{b}q)}\gamma^{\mu}\overset\leftrightarrow{\partial}_{\mu}\bar{H}^{(b\bar{q})}]+\text{H.c.},\\
\mathcal{L}_{D}=&\ ig_{D}\text{Tr}[D_{\mu\nu}^{(b\bar{b})}\bar{H}^{(\bar{b}q)}\overset\leftrightarrow{\partial}_{\mu}\gamma^{\nu}\bar{H}^{(b\bar{q})}]+\text{H.c.},\\
\mathcal{L}_{\mathbb{P}}=&\
ig_{H}\text{Tr}[H^{(b\bar{q})\text{j}}\gamma_{\mu}\gamma_{5}(\mathcal{A}^{\mu})^{\text{i}}_{\text{j}}\bar{H}_{\text{i}}^{(b\bar{q})}]
\label{eq:Lagrangians}
\end{split}
\end{equation}
with $\overset\leftrightarrow{\partial}\equiv\overset\rightarrow{\partial}-\overset\leftarrow{\partial}$. Here, the abbreviations $S^{(b\bar{b})}$ and $D_{\mu\nu}^{(b\bar{b})}$ represent the $S$-wave and $D$-wave multiplets of bottomonium, respectively, {\it i.e.},
\begin{equation}
S^{(b\bar{b})}=\frac{1+\slashed{\upsilon}}{2}\left[\Upsilon^{\mu}\gamma_{\mu}-\eta_{b}\gamma_{5}\right]\frac{1-\slashed{\upsilon}}{2},
\end{equation}
\begin{equation}
\begin{split}
D^{(b\bar{b})\mu\nu}=&\frac{1+\slashed{\upsilon}}{2}\left[\Upsilon_{3}^{\mu\nu\alpha}\gamma_{\alpha}+\frac{1}{\sqrt{6}}\left(\varepsilon^{\mu\alpha\beta\rho}\upsilon_{\alpha}\gamma_{\beta}\Upsilon_{2\rho}^{\nu}+\varepsilon^{\nu\alpha\beta\rho}\upsilon_{\alpha}\gamma_{\beta}\right.\right.\\
&\left.\left.\times\Upsilon_{2\rho}^{\mu}\right)+\frac{\sqrt{15}}{10}\left[\left(\gamma^{\mu}-\upsilon^{\mu}\right)\Upsilon_{1}^{\nu}+\left(\gamma^{\nu}-\upsilon^{\nu}\right)\Upsilon_{1}^{\mu}\right]\right.\\
&\left.-\frac{1}{\sqrt{15}}\left(g^{\mu\nu}-\upsilon^{\mu}\upsilon^{\nu}\right)\gamma_{\alpha}\Upsilon_{1}^{\alpha}+\eta_{b2}^{\mu\nu}\gamma_{5}\right]\frac{1-\slashed{\upsilon}}{2}.
\end{split}
\end{equation}
Additionally, the $(0^-,1^-)$ doublets of bottom and antibottom mesons is abbreviated as  $H^{(b\bar{q})}$ and $H^{(\bar{b}q)}$, respectively, which can be expressed as
\begin{equation}
\begin{split}
H^{(b\bar{q})}=&\frac{1+\slashed{\upsilon}}{2}\left(\mathcal{B}^{*\mu}\gamma_{\mu}+i\mathcal{B}\gamma_{5}\right),\\
H^{(\bar{b}q)}=&\left(\bar{\mathcal{B}}^{*\mu}\gamma_{\mu}+i\bar{\mathcal{B}}\gamma_{5}\right)\frac{1-\slashed{\upsilon}}{2},
\end{split}
\end{equation}
where the normalization factor $\sqrt{m_{\mathcal{B}^{(*)}}}$ is neglected here. The $\bar{H}^{(b\bar{q})}$ and $\bar{H}^{(\bar{b}q)}$ fields can be obtained through $\bar{H}^{(b\bar{q})}=\gamma_{0}H^{\dagger(b\bar{q})}\gamma_{0}$ and $\bar{H}^{(\bar{b}q)}=\gamma_{0}H^{\dagger(\bar{b}q)}\gamma_{0}$. $\mathcal{A^{\mu}}$, the axial vector current of Nambu-Goldstone fields, is expressed as $\mathcal{A}^{\mu}=(\xi^{\dagger}\partial^{\mu}\xi-\xi\partial^{\mu}\xi^{\dagger})/2$ with $\xi=e^{i\mathbb{P}/f_{\pi}}$, where the pseudoscalar octet $\mathbb{P}$ is
\begin{equation}
\begin{split}
\mathbb{P}&=
\begin{pmatrix}
-\frac{\pi^0}{\sqrt{2}}+\frac{\eta_8}{\sqrt{6}}&\pi^{+}&K^{+}\\
-\pi^{-}&\frac{\pi^0}{\sqrt{2}}+\frac{\eta_8}{\sqrt{6}}&K^{0}\\
-K^{-}&\bar{K}^{0}&-\frac{2}{\sqrt{6}}\eta_8
\end{pmatrix}.\\
\end{split}
\end{equation}

With the above preparation, we expand the compact Lagrangians in Eq.~\eqref{eq:Lagrangians} to get the following effective Lagrangians
\begin{equation}
\begin{split}
\mathcal{L}_{\Upsilon\mathcal{B}^{(*)}\mathcal{B}^{(*)}}=&\
ig_{\Upsilon\mathcal{B}\mathcal{B}}\Upsilon^{\mu}(\partial_{\mu}\mathcal{B}^{\dagger}\mathcal{B}-\mathcal{B}^{\dagger}\partial_{\mu}\mathcal{B})\\
&+g_{\Upsilon\mathcal{B}\mathcal{B}^{*}}\varepsilon_{\mu\nu\alpha\beta}\partial^{\mu}\Upsilon^{\nu}(\mathcal{B}^{*\alpha\dagger}\overset\leftrightarrow{\partial^\beta}\mathcal{B}-\mathcal{B}^{\dagger}\overset\leftrightarrow{\partial^\beta}\mathcal{B}^{*\alpha})\\
&+ig_{\Upsilon\mathcal{B}^{*}\mathcal{B}^{*}}\Upsilon^{\mu}(\partial_{\nu}\mathcal{B}_{\mu}^{*\dagger}\mathcal{B}^{*\nu}-\mathcal{B}^{*\nu\dagger}\partial_{\nu}\mathcal{B}_{\mu}^{*}+\mathcal{B}^{*\nu\dagger}\overset\leftrightarrow\partial_{\mu}\mathcal{B}_{\nu}^{*}),
\label{eq:SBB}
\end{split}
\end{equation}
\begin{equation}
\begin{split}
\mathcal{L}_{\Upsilon_{1}\mathcal{B}^{(*)}\mathcal{B}^{(*)}}=&\
ig_{\Upsilon_{1}\mathcal{B}\mathcal{B}}\Upsilon_{1}^{\mu}(\partial_{\mu}\mathcal{B}^{\dagger}\mathcal{B}-\mathcal{B}^{\dagger}\partial_{\mu}\mathcal{B})\\
&{+g_{\Upsilon_{1}\mathcal{B}\mathcal{B}^{*}}\varepsilon_{\mu\nu\alpha\beta}\partial^{\mu}\Upsilon_{1}^{\nu}(\mathcal{B}^{*\alpha\dagger}\overset\leftrightarrow{\partial^\beta}\mathcal{B}-\mathcal{B}^\dagger\overset\leftrightarrow{\partial^\beta}\mathcal{B}^{*\alpha})}\\
&+ig_{\Upsilon_{1}\mathcal{B}^{*}\mathcal{B}^{*}}\Upsilon_{1}^{\mu}(\partial_{\nu}\mathcal{B}_{\mu}^{*\dagger}\mathcal{B}^{*\nu}-\mathcal{B}^{*\nu\dagger}\partial_{\nu}\mathcal{B}_{\mu}^{*}\\
&+4\mathcal{B}^{*\nu\dagger}\overset\leftrightarrow\partial_{\mu}\mathcal{B}_{\nu}^{*}),
\label{eq:D1BB}
\end{split}
\end{equation}
\begin{equation}
\begin{split}
\mathcal{L}_{\Upsilon_2\mathcal{B}^{(*)}\mathcal{B}^{*}}=&
-ig_{\Upsilon_2\mathcal{B}\mathcal{B}^*}\Upsilon_2^{\mu\nu}(\mathcal{B}^\dagger\overset\leftrightarrow{\partial_{\mu}}\mathcal{B}^*_{\nu}-\mathcal{B}^{*\dagger}_{\nu}\overset\leftrightarrow{\partial_{\mu}}\mathcal{B})\\
&+g_{\Upsilon_2\mathcal{B}^*\mathcal{B}^*}\varepsilon_{\mu\nu\alpha\beta}\partial^{\nu}\Upsilon_2^{\rho\beta}
(\mathcal{B}^{*\dagger}_\rho\overset\leftrightarrow{\partial^{\mu}}\mathcal{B}^{*\alpha}+\mathcal{B}^{*\alpha\dagger}\overset\leftrightarrow{\partial^{\mu}}\mathcal{B}^{*}_\rho),
\label{eq:D2BB}
\end{split}
\end{equation}
\begin{equation}
\mathcal{L}_{\Upsilon_3\mathcal{B}^*\mathcal{B}^*}=
-ig_{\Upsilon_3\mathcal{B}^*\mathcal{B}^*}\Upsilon_3^{\mu\nu\alpha}(\mathcal{B}^{*\dagger}_\nu\overset\leftrightarrow{\partial_{\mu}}\mathcal{B}^*_\alpha+\mathcal{B}^{*\dagger}_\alpha\overset\leftrightarrow{\partial_{\mu}}\mathcal{B}^*_\nu),
\label{eq:D3BB}
\end{equation}
\begin{equation}
\begin{split}
\mathcal{L}_{\mathcal{B}^{(*)}\mathcal{B}^{*}\mathbb{P}}=&\ ig_{\mathcal{B}\mathcal{B}^{*}\mathbb{P}}(\mathcal{B}_{\mu}^{*\dagger}\mathcal{B}-\mathcal{B}^{\dagger}\mathcal{B}_{\mu}^{*})\partial^{\mu}\mathbb{P}\\
&-g_{\mathcal{B}^{*}\mathcal{B}^{*}\mathbb{P}}\varepsilon_{\mu\nu\alpha\beta}\partial^{\mu}\mathcal{B}^{*\dagger\nu}\partial^{\alpha}\mathcal{B}^{*\beta}\mathbb{P},
\label{eq:PseudoscalarBB}
\end{split}
\end{equation}
where $\mathcal{B}^{(*)\dagger}$ and $\mathcal{B}^{(*)}$ are defined as $\mathcal{B}^{(*)\dagger} = (B^{(*)+}, B^{(*)0}, B_{s}^{(*)0})$ and $\mathcal{B}^{(*)} = (B^{(*)-}, \bar{B}^{(*)0}, \bar{B}_{s}^{(*)0})^{T}$, respectively.

Now, we can write down the concrete amplitudes according to the diagrams in Fig.~\ref{fig:DJeta} by the Feynman rules listed in Appendix~\ref{app01}. With the first diagram in Fig.~\ref{fig:DJeta} as an example, the expression of its amplitude is
\begin{equation}
\begin{split}
\mathcal{M}_{4S}^{(1-1)}=
&\frac{|\vec{p}_\eta|}{32\pi^2m}\int d\Omega\
\epsilon_{\Upsilon^\prime}^{\mu}g_{\Upsilon^\prime BB}(q_{1\mu}-q_{2\mu})\\
&\times g_{BB^*\eta}p_1^\alpha g_{BB^*\Upsilon_1}\varepsilon_{\kappa\lambda\xi\tau}p_2^\lambda
\varepsilon_{\Upsilon_1}^{*\xi}(q^{\kappa}-q_2^{\kappa})\\
&\times \frac{-g_{\alpha}^{\tau}+q_{\alpha}q^{\tau}/m_{B^{*}}^2}{q^2-m_{B^{*}}^2}
\mathcal{F}^{2}(q^2,m_{B^{*}}^2)
\end{split}
\end{equation}
based on the Cutkosky cutting rule. And then, the remaining amplitudes can be obtained similarly.

Under the $4S$-$3D$ mixing scheme, the total amplitude is
\begin{equation}
\mathcal{M}_J^{\text{Total}}=4\sum_{\text{i}=1}^{\text{i}_{\text{max}}}\mathcal{M}_{4S}^{\text{(J-i)}}\sin{\theta}
+4\sum_{\text{j}=1}^{\text{j}_{\text{max}}}\mathcal{M}_{3D}^{\text{(J-j)}}\cos{\theta},
\end{equation}
where the superscript $\text{i(j)}$ denotes the $\text{i(j)}$-th amplitudes from the bottom meson loops in the above diagrams, the index $J$ denotes differential final $D$-wave bottomonium states $\Upsilon(1^3D_J)$, and the subscripts $4S$ and $3D$ is applied to distinguish the contributions from the $\Upsilon(4S)$ and $\Upsilon(3D)$ components, respectively. The mixing angle $\theta\approx33\degree$ is suggested in Refs.~\cite{Li:2021jjt,Bai:2022cfz}. In addition, the charge conjugation transformation ($B^{(*)}\leftrightarrow\bar{B}^{(*)}$) and the isospin transformations on the bridged $B^{(*)}$ mesons ($B^{(*)0}\leftrightarrow B^{(*)+}$ and $\bar{B}^{(*)0}\leftrightarrow B^{(*)-}$) require a fourfold factor.

Finally, the decay widths of the transitions of the $\Upsilon(10753)$ into a low-lying $D$-wave bottomonium by emitting a light pseudoscalar meson $\eta$ can be evaluated by
\begin{equation}
\Gamma[\Upsilon(10753)\to\Upsilon(1^3D_J)\eta]=\frac{1}{3}\frac{|\vec{p}_\eta|}{8\pi m^{2}}|\overline{\mathcal{M}_{J}^{\text{Total}}}|^{2},
\end{equation}
where the overbar above amplitude denotes the sum over the polarizations of the $\Upsilon(1^3D_J)$. The coefficient $1/3$ comes from averaging over spins of the initial state. Besides, $m$ is the mass of the $\Upsilon(10753)$, and $\vec{p}_\eta$ is the three-momentum of $\eta$ meson in the rest frame of the initial  $\Upsilon(10753)$.

\section{Numerical result}
\label{sec03}

Before displaying the numerical results, we need to introduce how to fix the values of these involved parameters, which include the masses and the related coupling constants. For the mass and width of the $\Upsilon(10753)$, the measured central values from the Belle Collaboration,  $m_{\Upsilon(10753)}=10.753$ GeV and $\Gamma_{\Upsilon(10753)}=35.5$ MeV~\cite{Belle:2019cbt}, are adopted in our calculation. For the mass of the $\Upsilon(1^3D_2)$, we take its experimental result $m_{\Upsilon(1^3D_2)}=10.164$ GeV \cite{BaBar:2010tqb}. For the masses of the $\Upsilon(1^3D_1)$ and $\Upsilon(1^3D_3)$ still missing in experiment, the theoretical results $10.153$ and $10.170$ GeV predicted in Ref.~\cite{Wang:2018rjg} are taken in our calculation, respectively. Moreover, the PDG values \cite{ParticleDataGroup:2020ssz} are used for other involved bottom mesons and $\eta$ meson in this work.

In the following, we should determine the relevant coupling constants. The coupling constants $g_{\Upsilon_1^\prime B^{(*)}B^{(*)}}$ depicting the coupling between the $\Upsilon(3D)$ and a pair of bottom mesons are extracted from the corresponding decay widths given in  Ref.~\cite{Wang:2018rjg}.
The corresponding coupling constants are listed in Table~\ref{tab:couplingconstants}.
And then, the $g_{\Upsilon^\prime BB}$ value is determined by the corresponding partial decay width given in Ref.~\cite{Wang:2018rjg}, while the $g_{\Upsilon^\prime BB^{*}}$ value can be fixed by $g_{\Upsilon^\prime BB}$ and the relations shown in Eq.~\eqref{eq:relations} which is from the heavy quark symmetry. For convenience of reader,
the values of these coupling constants are also collected into  Table~\ref{tab:couplingconstants}.

\begin{table}[htbp]
\centering
\caption{The values of these involved coupling constants in our calculation.}
\label{tab:couplingconstants}
\renewcommand\arraystretch{1.05}
\begin{tabular*}{86mm}{c@{\extracolsep{\fill}}ccc}
\toprule[1pt]
\toprule[0.5pt]
$g$                   &$B\bar{B}$    &$B\bar{B}^{*}+\text{c.c}$    &$B^{*}\bar{B}^{*}$ \\
\midrule[0.5pt]
$\Upsilon(4S)$        &$13.22$      &$1.251\ \text{GeV}^{-1}$     &-- \\
$\Upsilon(3D)$        &$3.480$       &$0.393\ \text{GeV}^{-1}$     &$4.210$ \\
$\Upsilon(1^3D_1)$    &$427.0$     &$21.12\ \text{GeV}^{-1}$    &$43.06$ \\
$\Upsilon(1^3D_2)$    &--            &$407.0$                    &$13.41\ \text{GeV}^{-1}$ \\
$\Upsilon(1^3D_3)$    &--            &--                           &$333.8$ \\
$\eta$                &--            &$17.77$    &$3.351\ \text{GeV}^{-1}$ \\
\bottomrule[0.5pt]
\bottomrule[1pt]
\end{tabular*}
\end{table}

The coupling constants defined in Eq.~\eqref{eq:D1BB}, Eq.~\eqref{eq:D2BB}, and Eq.~\eqref{eq:D3BB} read as
\begin{equation}
\begin{split}
g_{\Upsilon_1BB}&=\frac{10}{\sqrt{15}}g_Dm_{B}\sqrt{m_{\Upsilon_1}},\\
g_{\Upsilon_1BB^*}&=\frac{5}{\sqrt{15}}g_D\sqrt{m_{B}m_{B^*}/m_{\Upsilon_1}},\\
g_{\Upsilon_1B^*B^*}&=\frac{1}{\sqrt{15}}g_Dm_{B^*}\sqrt{m_{\Upsilon_1}},\\
g_{\Upsilon_2BB^*}&=\sqrt{6}g_D\sqrt{m_{B}m_{B^*}m_{\Upsilon_2}},\\
g_{\Upsilon_2B^*B^*}&=\frac{2}{\sqrt{6}}g_Dm_{B^*}/\sqrt{m_{\Upsilon_2}},\\
g_{\Upsilon_3B^*B^*}&=2g_Dm_{B^*}\sqrt{m_{\Upsilon_3}},
\label{eq:relations}
\end{split}
\end{equation}
where $g_D=9.83\ \text{GeV}^{-3/2}$~\cite{Wang:2016qmz,Huang:2018cco}.

According to the SU(3) quark model, the observed $\eta$ and $\eta^\prime$ are mixing of the singlet $\eta_1$ and octet $\eta_8$,
\begin{equation}
\left(\begin{array}{c}
\eta\\
\eta^\prime
\end{array}\right)
=\left(\begin{array}{cc}
\cos{\theta_\eta} & -\sin{\theta_\eta}\\
\sin{\theta_\eta} & \cos{\theta_\eta}
\end{array}\right)
\left(\begin{array}{c}
\eta_8\\
\eta_1
\end{array}\right).
\end{equation}
Thus, the coupling constant $g_{B^{(*)}B^{(*)}\eta}$ can be expressed by the coupling constant $g_{H}$, {\it i.e.},
\begin{equation}
\frac{g_{BB^{*}\eta}}{\sqrt{m_{B}m_{B^{*}}}}=g_{B^{*}B^{*}\eta}=\frac{2g_{H}}{f_{\pi}}\frac{\cos\theta_\eta}{\sqrt{6}},
\end{equation}
where $g_{H}=0.569$ and $f_{\pi}=$ 131 MeV~\cite{Wang:2016qmz,Huang:2018cco,Huang:2018pmk}. The mixing angle $\theta=-19.1\degree$ had been fixed by the DM2 Collaboration~\cite{DM2:1988bfq}. We also collect the involved coupling constants in Table~\ref{tab:couplingconstants}.

\begin{figure}[htbp]\centering
  \includegraphics[width=86mm]{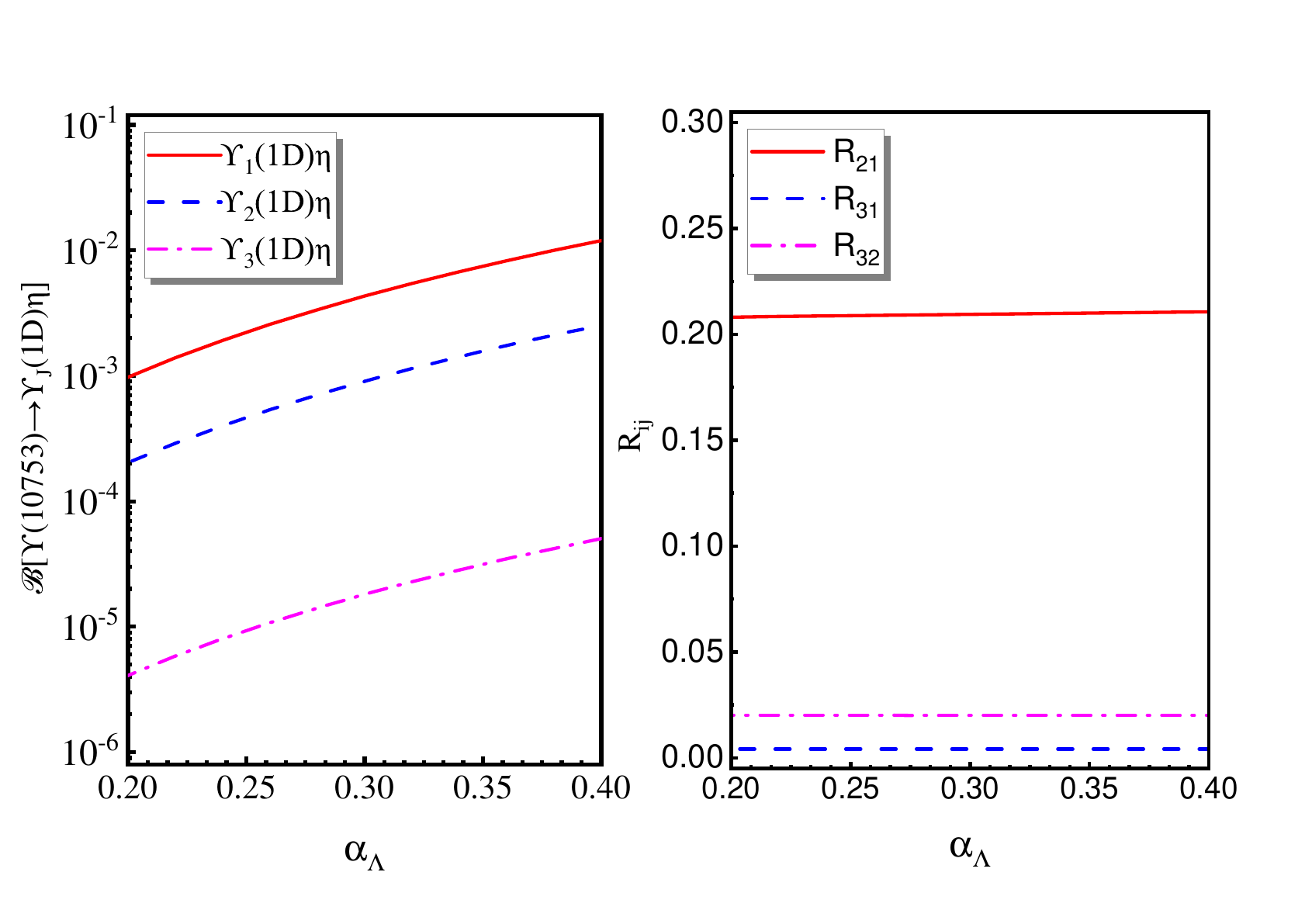}\\
  \caption{The $\alpha_{\Lambda}$ dependence of the calculated branching ratios $\mathcal{B}[\Upsilon(10753)\to\Upsilon(1^3D_J)\eta]$ ($J=1,2,3$) (left panel) and the ratios $R_{ij}=\mathcal{B}[\Upsilon(10753)\to\Upsilon(1^3D_i)\eta]/\mathcal{B}[\Upsilon(10753)\to\Upsilon(1^3D_j)\eta]$ (right panel).}
\label{fig:result}
\end{figure}

Until now, we have obtained all coupling constants involving in our calculation. However, there exists the phenomenological parameter $\alpha_\Lambda$ introduced in Eq.~\eqref{eq:formfactor} to parametrize the cutoff $\Lambda$. Since the cutoff $\Lambda$ should not deviate from the physical mass of the exchanged meson, $\alpha_\Lambda$ is restricted to be of the order of unity~\cite{Cheng:2004ru}. 
{Since there does not exist direct experimental data to constrain the $\alpha_{\Lambda}$ value, we have to borrow the  experience of the $\Upsilon(10860)$ transition into $\Upsilon_J(1D)\eta$ \cite{Belle:2018hjt}\footnote{As higher states above the $B\bar{B}$ threshold, the $\Upsilon(10753)$ is close to the $\Upsilon(10860)$, where these two states have similar widths \cite{ParticleDataGroup:2020ssz}.}, where the branching ratio of $\Upsilon(10860)\to \Upsilon_J(1D)\eta$ is of the order of magnitude of $10^{-3}$. 
To reach up to this order of magnitude, 
we should take the range $0.2\leq\alpha_\Lambda\leq0.4$, which 
satisfies the requirement of $\alpha_\Lambda$~\cite{Cheng:2004ru}.} 
For the $\Upsilon(10753)\to\Upsilon(1^3D_J)\eta$ ($J=1,2,3$) processes, the $\alpha_{\Lambda}$ dependence of the discussed branching ratios is displayed in left panel of Fig.~\ref{fig:result}.

From Fig.~\ref{fig:result}, we can summarize the behavior of the obtained branching ratios
\begin{eqnarray*}
\mathcal{B}[\Upsilon(10753)\to\Upsilon(1^3D_1)\eta]&=&(0.98 - 12.0)\times 10^{-3},\\
\mathcal{B}[\Upsilon(10753)\to\Upsilon(1^3D_2)\eta]&=&(0.20 - 2.52)\times 10^{-3},\\
\mathcal{B}[\Upsilon(10753)\to\Upsilon(1^3D_3)\eta]&=&(0.41 - 5.03)\times 10^{-5},
\end{eqnarray*}
by which the corresponding decay widths can be further presented as
\begin{eqnarray*}
\Gamma[\Upsilon(10753)\to\Upsilon(1^3D_1)\eta]&=&(34.7 - 425.3)\ \text{keV},\\
\Gamma[\Upsilon(10753)\to\Upsilon(1^3D_2)\eta]&=&(7.21 - 89.6)\ \text{keV},\\
\Gamma[\Upsilon(10753)\to\Upsilon(1^3D_3)\eta]&=&(0.15 - 1.79)\ \text{keV}.
\end{eqnarray*}

Additionally, we also notice that the ratios $R_{ij}=\mathcal{B}[\Upsilon(10753)\to\Upsilon(1^3D_i)\eta]/\mathcal{B}[\Upsilon(10753)\to\Upsilon(1^3D_j)\eta]$ (see the right panel of Fig.~\ref{fig:result}) act weakly dependence of $\alpha_{\Lambda}$, i.e.,
\begin{eqnarray*}
R_{21}&=&\frac{\mathcal{B}[\Upsilon(10753)\to\Upsilon(1^3D_2)\eta]}{\mathcal{B}[\Upsilon(10753)\to\Upsilon(1^3D_1)\eta]}\approx0.21,\\
R_{31}&=&\frac{\mathcal{B}[\Upsilon(10753)\to\Upsilon(1^3D_3)\eta]}{\mathcal{B}[\Upsilon(10753)\to\Upsilon(1^3D_1)\eta]}\approx0.004,\\
R_{32}&=&\frac{\mathcal{B}[\Upsilon(10753)\to\Upsilon(1^3D_3)\eta]}{\mathcal{B}[\Upsilon(10753)\to\Upsilon(1^3D_2)\eta]}\approx0.02.
\end{eqnarray*}
which show that the $\Upsilon(10753)\to\Upsilon(1^3D_3)\eta$ decay is suppressed compared with the $\Upsilon(10753)\to\Upsilon(1^3D_{1})\eta$
and $\Upsilon(10753)\to\Upsilon(1^3D_{2})\eta$ decays. Thus, it is difficult to observe the $\Upsilon(1^3D_3)$ mode via the $\Upsilon(10753)\to\Upsilon(1^3D_3)\eta$ decay. The sizable branching ratios of the $\Upsilon(10753)\to\Upsilon(1^3D_{1})\eta$
and $\Upsilon(10753)\to\Upsilon(1^3D_{2})\eta$ decays indicate the probability of finding out them in Belle II.
Thus, experimental search for them will be an interesting task for future experiment like Belle II.

\section{Discussion and conclusion}
\label{sec04}

As a vector bottomonium candidate \cite{Wang:2018rjg}, the recently reported $\Upsilon(10753)$ by Belle exists  in the $e^+e^-\to \Upsilon(nS)\pi^+\pi^-$ ($n=1,2,3$) processes~\cite{Belle:2019cbt}. The $\Upsilon(10753)$ is a crucial state when constructing the bottomonium family. The study of its hidden-bottom decays is an important aspect to reflect the spectroscopy behavior of the $\Upsilon(10753)$ \cite{Li:2021jjt,Bai:2022cfz}. In this work, we calculate the $\Upsilon(10753)\to\Upsilon(1^3D_J)\eta$ decays, which are involved in the observed
$\Upsilon(1^3D_2)$ \cite{BaBar:2010tqb} and two missing bottomonia $\Upsilon(1^3D_1)$ and $\Upsilon(1^3D_3)$. Since the coupled-channel effect cannot be ignored for higher bottomonia \cite{Wang:2018rjg}, we should introduce the hadronic loop mechanism when exploring the decay behavior of the $\Upsilon(10753)\to\Upsilon(1^3D_J)\eta$ decays, and find that the $\Upsilon(10753)\to\Upsilon(1^3D_1)\eta$  and $\Upsilon(10753)\to\Upsilon(1^3D_2)\eta$ decay channels have sizable branching ratios. Thus, it is possible to find out these two predicted decay modes at Belle II. Different from these two decay channels, the
$\Upsilon(10753)\to\Upsilon(1^3D_3)\eta$ decay is suppressed. Thus, searching for $\Upsilon(10753)\to\Upsilon(1^3D_3)\eta$
is not promising only if more data are accumulated in experiments.

\begin{figure}[htbp]\centering
  \includegraphics[width=80mm]{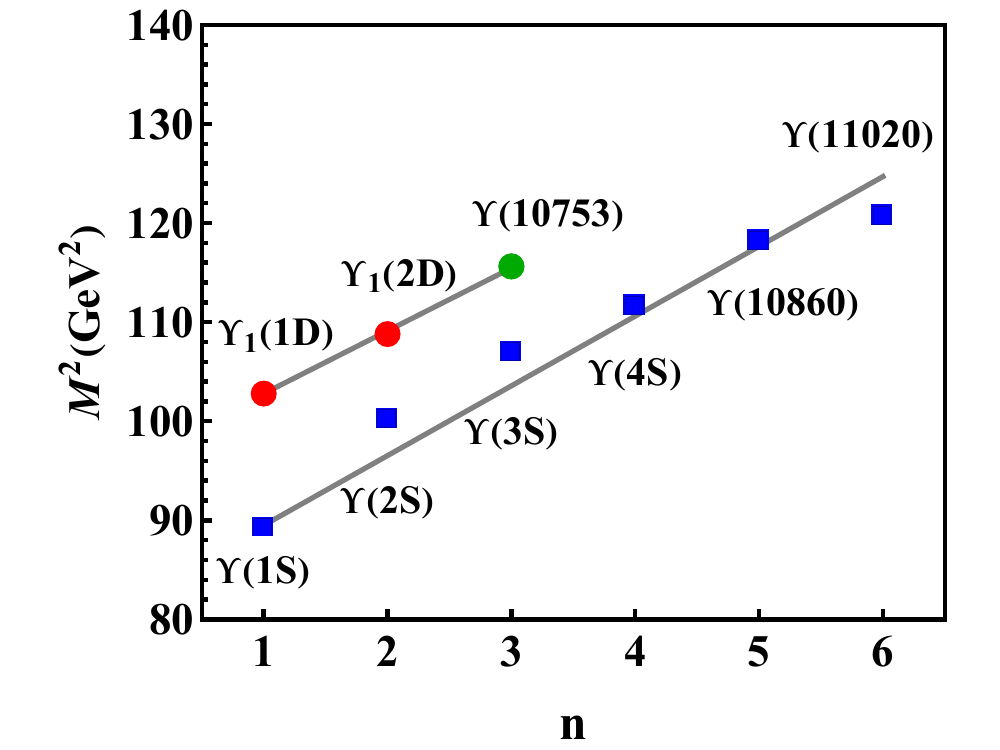}\\
  \caption{The Regge trajectories of the $\Upsilon(nS)$ and $\Upsilon_1(nD)$ series. Here, the blue squares denote the experimental data, the red cycles denote the averaged values of the predictions in Refs.~\cite{Godfrey:2015dia,Segovia:2016xqb,Wang:2018rjg,Deng:2016ktl} and the green one is the measured mass of the $\Upsilon(10753)$. The trajectory slopes are determined as $\mu^2=7.0403$ GeV$^2$ and $6.3364$ GeV$^2$  for the $\Upsilon(nS)$ and $\Upsilon_1(nD)$ series, respectively.}
\label{fig:Regge}
\end{figure}

In the following, we should discuss the mass spectrum of vector bottomonia.
As listed in PDG \cite{ParticleDataGroup:2020ssz}, there were
$\Upsilon(1S)$, $\Upsilon(1S)$, $\Upsilon(3S)$, $\Upsilon(4S)$, $\Upsilon(10860)$, and $\Upsilon(11020)$. We find that they form a Regge trajectory
as shown in Fig. \ref{fig:Regge}.
This Regge trajectory satisfies the relation $M^2=M_0^2+(n-1)\mu^2$ \cite{Regge:1959mz,Regge:1960zc,Chew:1961ev,Chew:1962eu,Collins:1971ff,Anisovich:2000kxa,Guo:2019wpx,Guo:2022xqu}. Here, $M_0$ is the mass of the ground state, $M$ denotes the mass of the radial excitation state with the radial quantum number $n$, and $\mu^2=7.0403$ GeV$^2$ is the slope of the Regge trajectory.
Although the $\Upsilon(10753)$ is suggested as the mixture of $4S$ and $3D$ states of bottomonium, the $\Upsilon(10753)$ has main component of $3D$ state. Thus, we may take $\Upsilon(10753)$ and the predicted $\Upsilon(1^3D_1)$ and $\Upsilon(2^3D_1)$ ~\cite{Godfrey:2015dia,Segovia:2016xqb,Wang:2018rjg,Deng:2016ktl} to construct
another Regge trajectory (see Fig. \ref{fig:Regge}), which have slope $\mu^2=6.3364$ GeV$^2$. This slope is similar to that for the $S$-wave bottomonia.
Thus, searching for the missing $\Upsilon(1^3D_1)$ and $\Upsilon(2^3D_1)$ bottononia in future experiment will be helpful to test this Regge trajectory behavior. It is obvious that the present work  of studying the $\Upsilon(10753)\to\Upsilon(1^3D_1)\eta$ shows the potential of finding out the $\Upsilon(1^3D_1)$ bottomonium.

With running of Belle II, the physics relevant to the $\Upsilon(10753)$ should be paid more attention. Searching for different decay modes of the $\Upsilon(10753)$ is crucial step of establishing the $\Upsilon(10753)$ as bottomonium. We hope that the present work may provide valuable information to future experimental exploration.

\section*{ACKNOWLEDGMENTS}
This work is supported by the China National Funds for Distinguished Young Scientists under Grant No. 11825503, National Key Research and Development Program of China under Contract No. 2020YFA0406400, the 111 Project under Grant No. B20063, and the National Natural Science Foundation of China under Grant No. 12047501.

\appendix

\section{The Feynman rules for the interaction vertexes}
\label{app01}
In this appendix, the Feynman rules for the involved interaction vertexes are presented. The concrete information includes
\begin{eqnarray}
\raisebox{-15pt}{\includegraphics[width=0.16%
\textwidth]{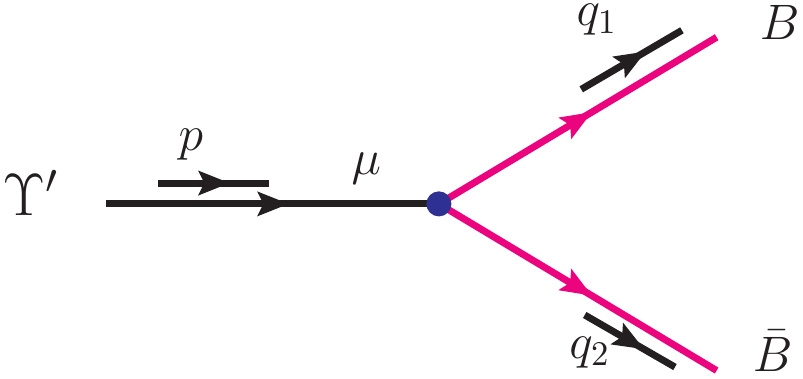}}           &\widehat{=}
&-g_{\Upsilon^\prime BB}\epsilon_{\Upsilon^\prime}^{\mu}(q_{1\mu}-q_{2\mu}),\\
\raisebox{-15pt}{\includegraphics[width=0.16%
\textwidth]{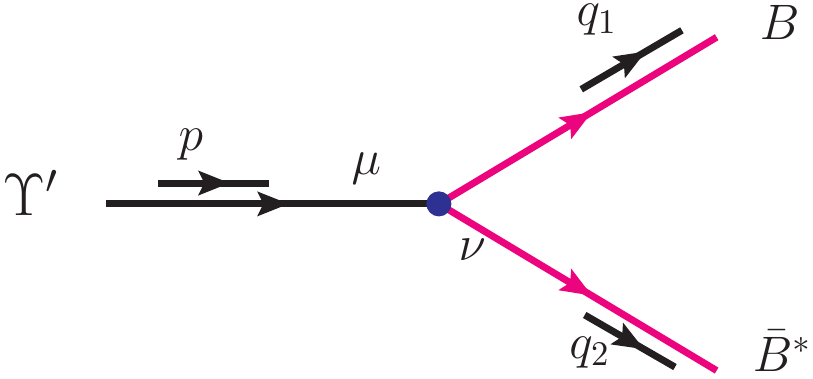}}       &\widehat{=}
&g_{\Upsilon^\prime BB^*}\varepsilon_{\alpha\beta\mu\nu}\epsilon_{\Upsilon^\prime}^{\mu}p^{\beta}(q_2^{\alpha}-q_1^{\alpha})\epsilon_{\bar{B}^{*}}^{*\nu},\\
\raisebox{-15pt}{\includegraphics[width=0.16%
\textwidth]{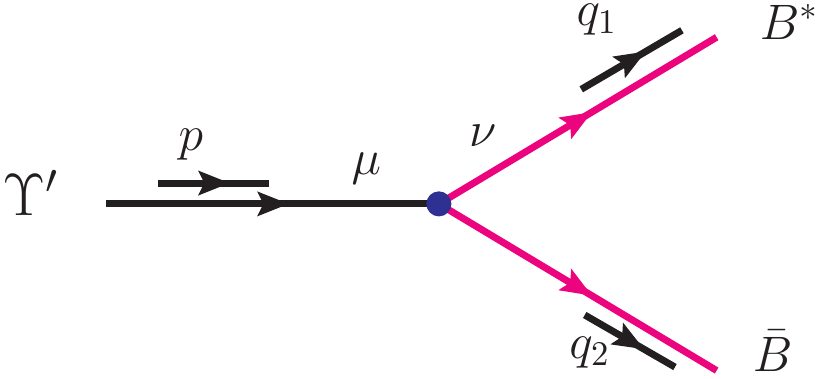}}       &\widehat{=}
&g_{\Upsilon^\prime BB^*}\varepsilon_{\alpha\beta\mu\nu}\epsilon_{\Upsilon^\prime}^{\mu}p^{\beta}(q_1^{\alpha}-q_2^{\alpha})\epsilon_{B^{*}}^{*\nu},\\
\raisebox{-15pt}{\includegraphics[width=0.16%
\textwidth]{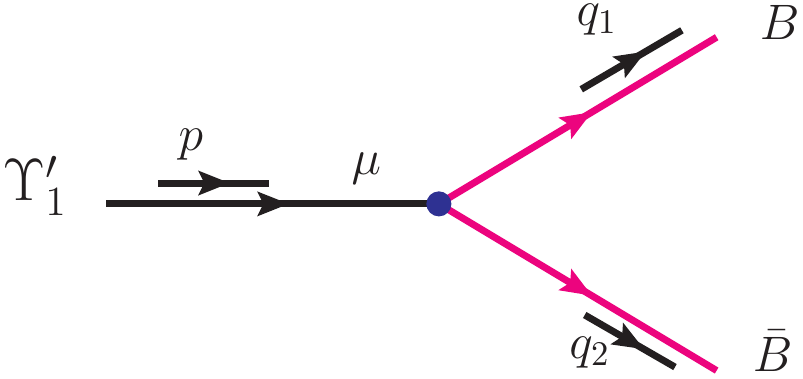}}           &\widehat{=}
&-g_{\Upsilon_1^\prime BB}\epsilon_{\Upsilon_1^\prime}^{\mu}(q_{1\mu}-q_{2\mu}),\\
\raisebox{-15pt}{\includegraphics[width=0.16%
\textwidth]{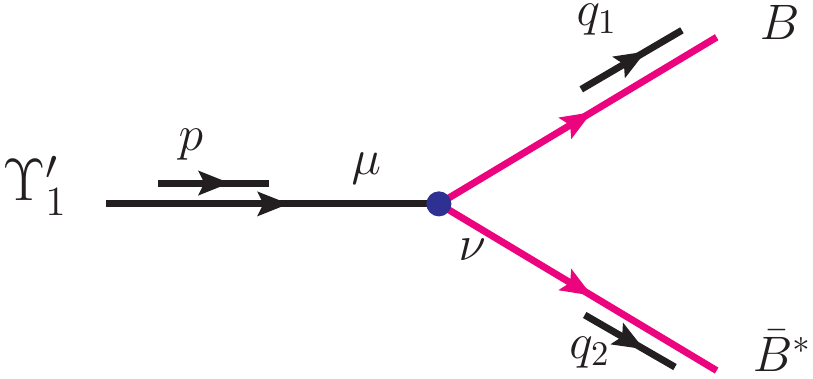}}       &\widehat{=}
&g_{\Upsilon_1^\prime BB^*}\varepsilon_{\alpha\beta\mu\nu}\epsilon_{\Upsilon_1^\prime}^{\mu}p^{\beta}(q_2^{\alpha}-q_1^{\alpha})\epsilon_{\bar{B}^{*}}^{*\nu},\\
\raisebox{-15pt}{\includegraphics[width=0.16%
\textwidth]{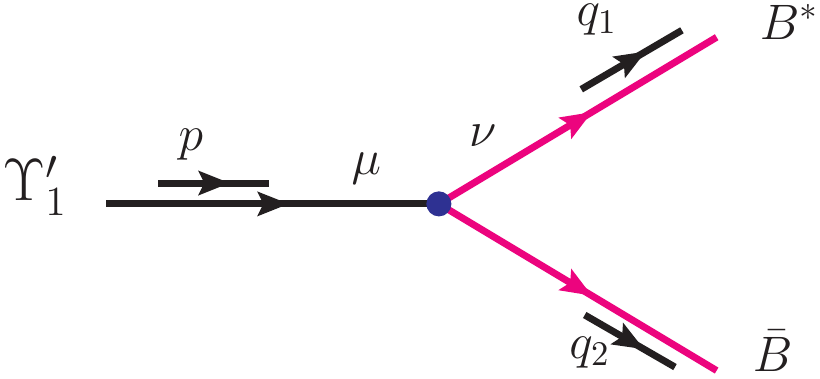}}       &\widehat{=}
&g_{\Upsilon_1^\prime BB^*}\varepsilon_{\alpha\beta\mu\nu}\epsilon_{\Upsilon_1^\prime}^{\mu}p^{\beta}(q_1^{\alpha}-q_2^{\alpha})\epsilon_{B^{*}}^{*\nu},\\
\raisebox{-15pt}{\includegraphics[width=0.16%
\textwidth]{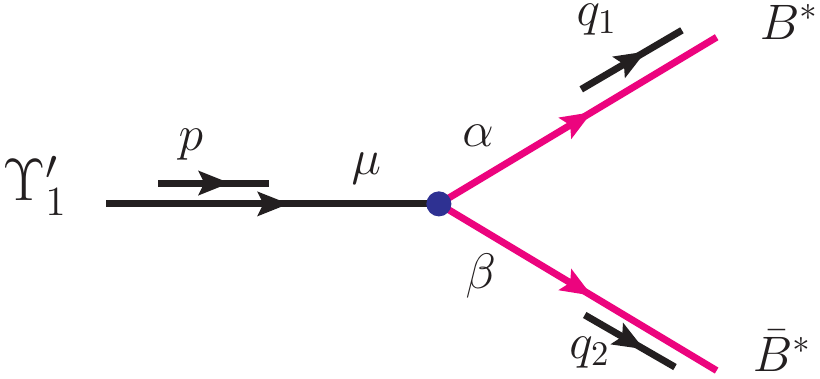}}   &\widehat{=}
&\begin{split}
&-g_{\Upsilon_1^\prime B^*B^*}\epsilon_{\Upsilon_1^\prime}^{\mu}(-4g_{\alpha\beta}(q_{1\mu}-q_{2\mu})\\
&+g_{\alpha\mu}q_{1\beta}-g_{\beta\mu}q_{2\alpha})\epsilon_{B^{*}}^{*\alpha}\epsilon_{\bar{B}^{*}}^{*\beta},
\end{split}\\
\raisebox{-15pt}{\includegraphics[width=0.16%
\textwidth]{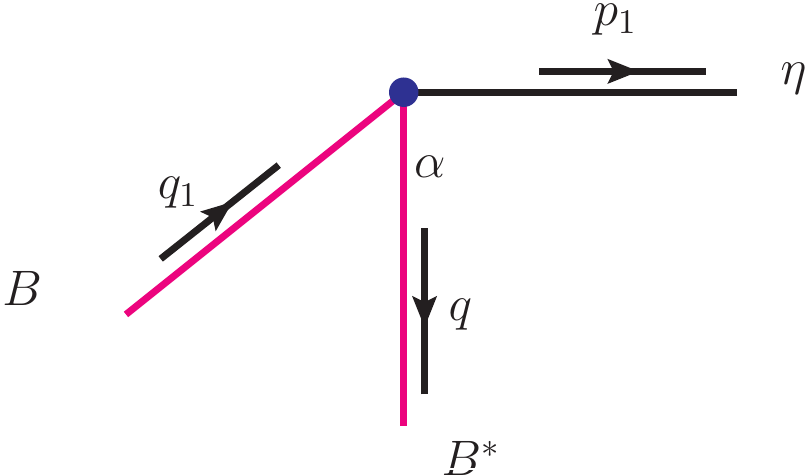}}           &\widehat{=}
&-g_{BB^*\eta^{(\prime)}}p_1^{\alpha}\epsilon_{B^{*}\alpha}^{*},\\
\raisebox{-15pt}{\includegraphics[width=0.16%
\textwidth]{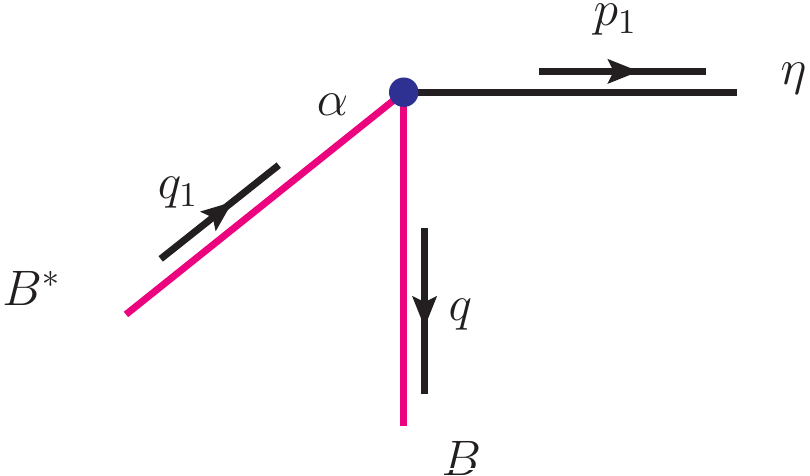}}       &\widehat{=}
&g_{BB^*\eta^{(\prime)}}p_1^{\alpha}\epsilon_{B^{*}\alpha},\\
\raisebox{-15pt}{\includegraphics[width=0.16%
\textwidth]{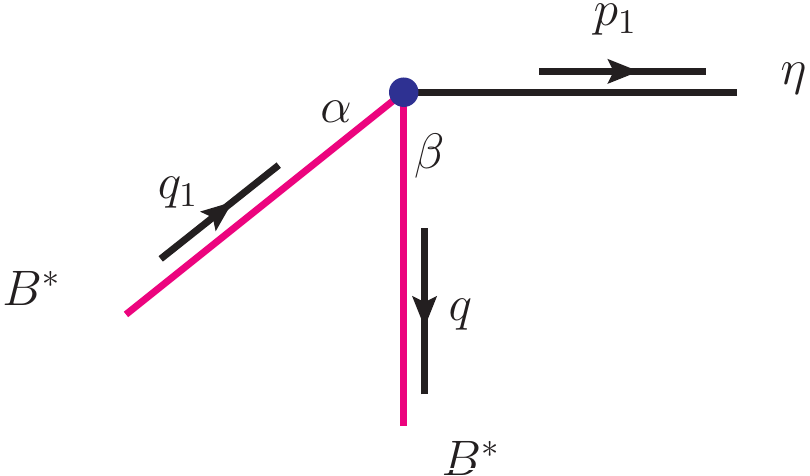}}       &\widehat{=}
&-g_{B^*B^*\eta^{(\prime)}}\varepsilon_{\mu\nu\alpha\beta}q_1^{\nu}q^{\mu}\epsilon_{B^{*}}^{\alpha}\epsilon_{B^{*}}^{*\beta}.\\
\raisebox{-15pt}{\includegraphics[width=0.16%
\textwidth]{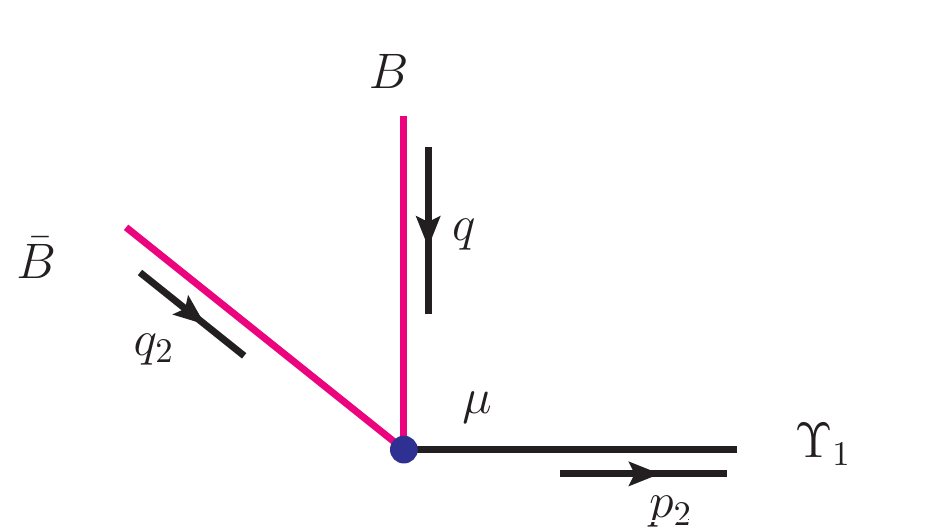}}           &\widehat{=}
&g_{\Upsilon_1 BB}\epsilon_{\Upsilon_1}^{*\mu}(q_{2\mu}-q{\mu}),\\
\raisebox{-15pt}{\includegraphics[width=0.16%
\textwidth]{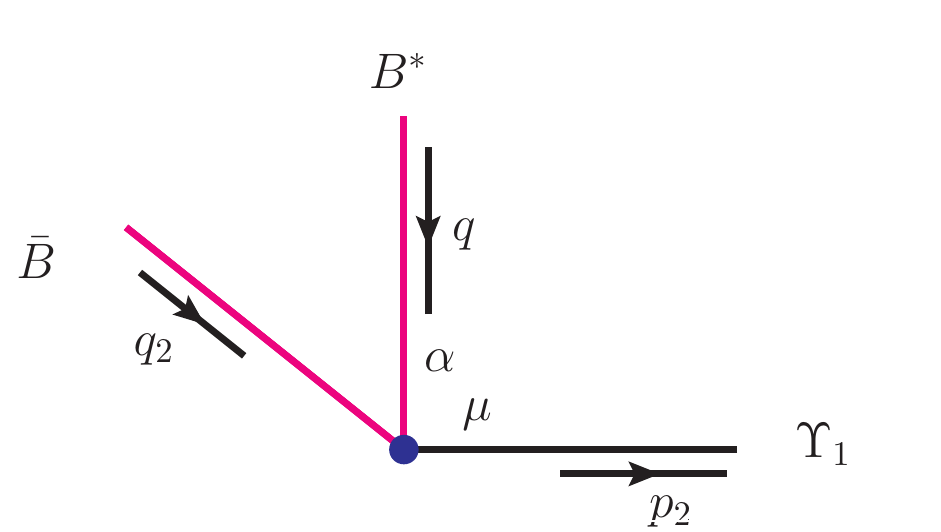}}       &\widehat{=}
&g_{\Upsilon_1 BB^*}\varepsilon_{\mu\nu\alpha\beta}p_2^{\nu}\epsilon_{\Upsilon_1}^{*\mu}\epsilon_{B^*}^{\alpha}(q^{\beta}-q_{2}^{\beta}),\\
\raisebox{-15pt}{\includegraphics[width=0.16%
\textwidth]{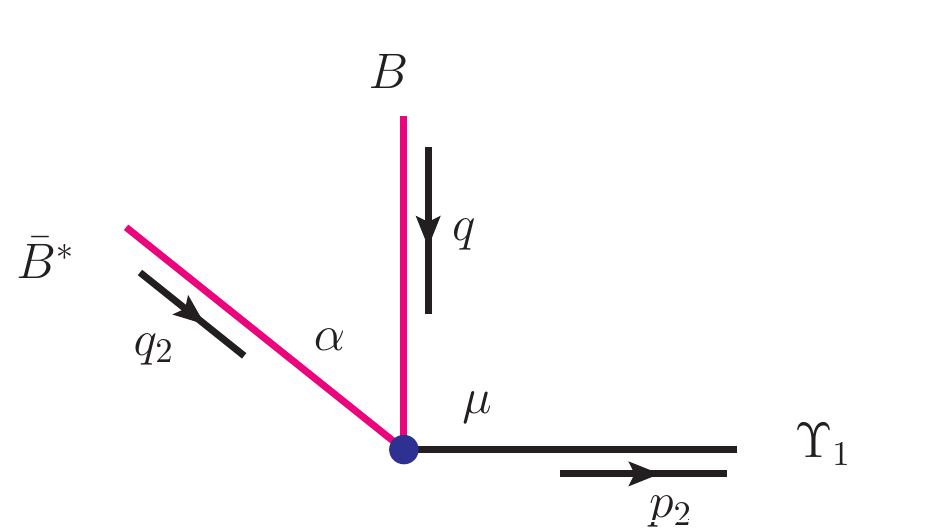}}       &\widehat{=}
&g_{\Upsilon_1 BB^*}\varepsilon_{\mu\nu\alpha\beta}p_2^{\nu}\epsilon_{\Upsilon_1}^{*\mu}\epsilon_{\bar{B}^*}^{\alpha}(q_2^{\beta}-q^{\beta}),\\
\raisebox{-15pt}{\includegraphics[width=0.16%
\textwidth]{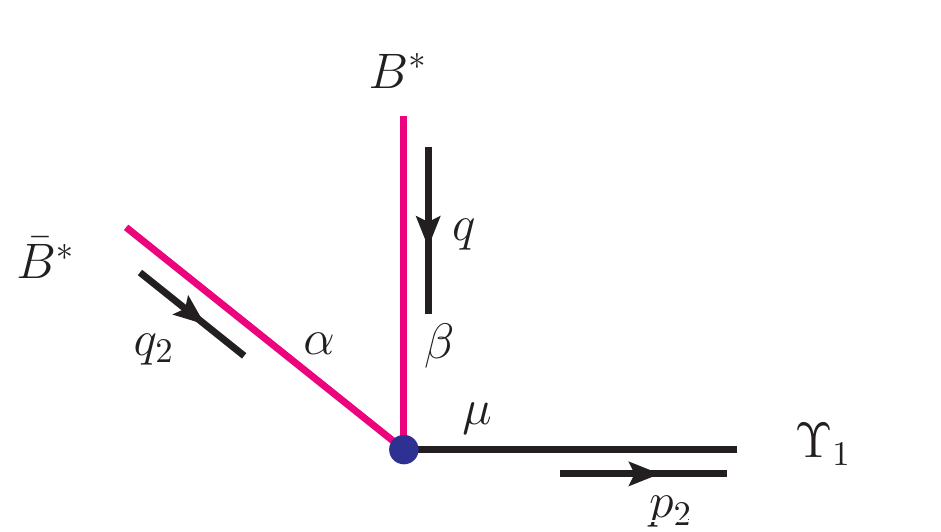}}       &\widehat{=}
&\begin{split}
&g_{\Upsilon_1B^*B^*}\epsilon_{\Upsilon_1}^{*\mu}\epsilon_{B^*}^{\beta}\epsilon_{\bar{B}^*}^{\alpha}(g_{\alpha\mu}q_{2\beta}\\
&-g_{\beta\mu}q_{\alpha}+4g_{\alpha\beta}(q_{\mu}-q_{2\mu})),
\end{split}\\
\raisebox{-15pt}{\includegraphics[width=0.16%
\textwidth]{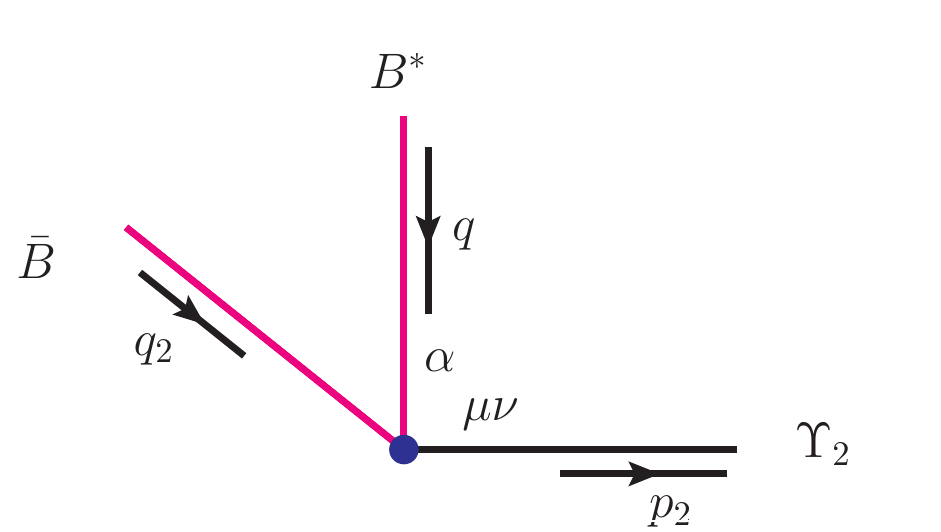}}       &\widehat{=}
&-g_{\Upsilon_2BB^*}\epsilon_{\Upsilon_2}^{*\mu\nu}g_{\alpha\nu}\epsilon_{B^*}^{\alpha}(q_{\mu}-q_{2\mu}),\\
\raisebox{-15pt}{\includegraphics[width=0.16%
\textwidth]{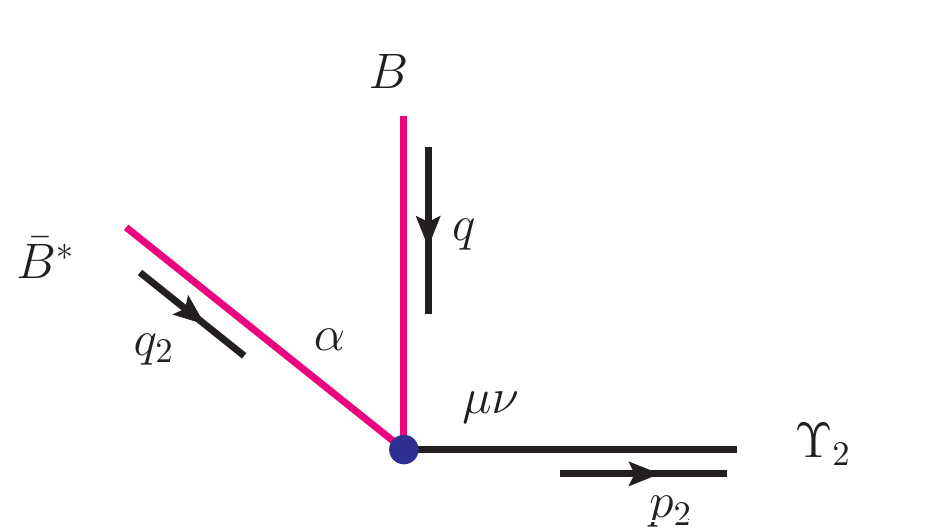}}       &\widehat{=}
&g_{\Upsilon_2BB^*}\epsilon_{\Upsilon_2}^{*\mu\nu}g_{\alpha\nu}\epsilon_{\bar{B}^*}^{\alpha}(q_{\mu}-q_{2\mu}),\\
\raisebox{-15pt}{\includegraphics[width=0.16%
\textwidth]{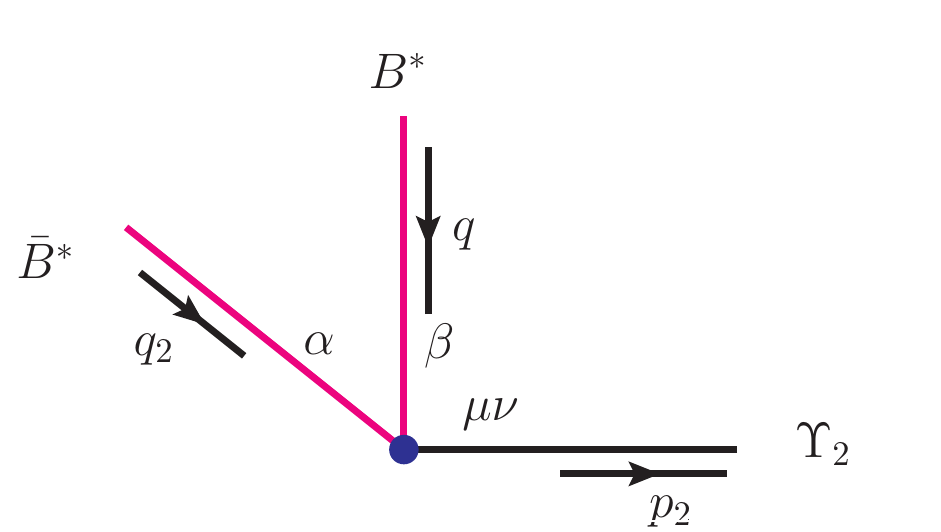}}       &\widehat{=}
&\begin{split}
&g_{\Upsilon_2B^*B^*}\varepsilon_{\kappa\lambda\xi\nu}p_{2}^{\lambda}\epsilon_{\Upsilon_2}^{*\mu\nu}(q^{\kappa}-q_{2}^{\kappa})\\
&\times(g_{\alpha\mu}g_{\beta}^{\xi}+g_{\alpha}^{\xi}g_{\beta\mu})\epsilon_{B^*}^{\beta}\epsilon_{\bar{B}^*}^{\alpha},
\end{split}\\
\raisebox{-15pt}{\includegraphics[width=0.16%
\textwidth]{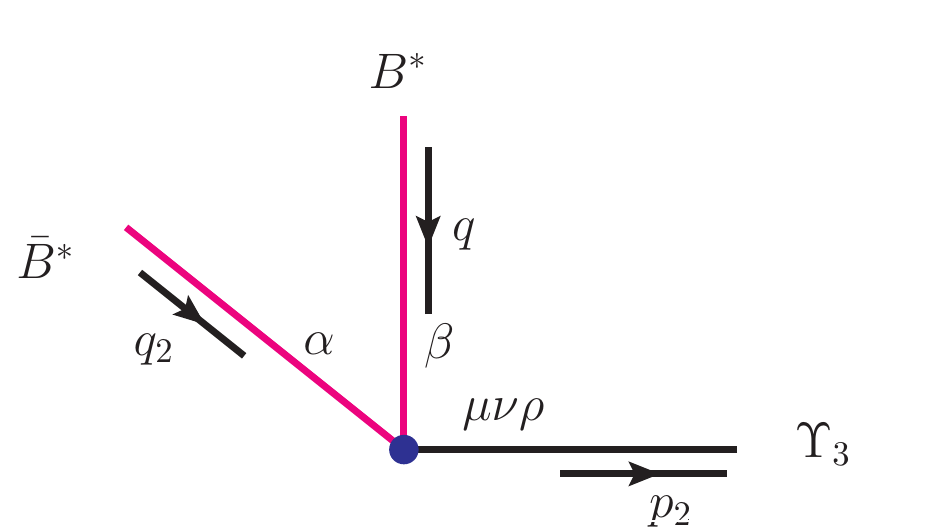}}       &\widehat{=}
&\begin{split}
&-g_{\Upsilon_3B^*B^*}\epsilon_{\Upsilon_3}^{*\mu\nu\rho}(g_{\alpha\nu}g_{\beta\rho}+g_{\alpha\rho}g_{\beta\nu})\\
&\times(q_{\mu}-q_{2\mu})\epsilon_{B^*}^{\beta}\epsilon_{\bar{B}^*}^{\alpha}.
\end{split}
\end{eqnarray}

\end{document}